\documentclass[final]{aipproc}
\layoutstyle{6x9}

\begin{document}

\title{System Size and Energy Dependence \\
 of Elliptic Flow}

\classification{25.75.-q, 25.75.Ld}
\keywords{relativistic heavy ion collisions, elliptic flow, Au+Au, Cu+Cu}

\author{Alice C. Mignerey for the PHOBOS Collaboration\footnotemark}{address={Department of Chemistry and Biochemistry, University of Maryland, College Park, MD 20742, USA}
}

\begin{abstract} The elliptic flow  $\mathrm{v_{2}}$ is presented for the Cu+Cu collisions at \mbox{$\mathrm{\sqrt{s_{NN}}}$} = 62.4 and 200 GeV, as a function of pseudorapidity.  Comparison to results for the Au+Au collisions at the same energies shows a reduction of about 20\% in the flow observed for a centrality selection of 0-40\%.  The centrality dependent flow, expressed as a function of the number of participants $\mathrm{N_{part}}$, is compared for the Cu+Cu and Au+Au systems using two definitions of eccentricity, the standard definition $\mathrm{\epsilon_{standard}}$ and a participant eccentricity $\mathrm{\epsilon_{part}}$. The $\mathrm{v_{2}}$/$\mathrm{\langle\epsilon_{part}\rangle}$ as a function of $\mathrm{N_{part}}$, for the Au+Au and Cu+Cu collisions are consistent within errors, while $\mathrm{v_{2}}$/$\mathrm{\langle\epsilon_{standard}\rangle}$ gives unrealistically large values for Cu+Cu, especially for central collision.
\end{abstract}
\maketitle \footnotetext{Collaboration: B.Alver$^4$, B.B.Back$^1$,
M.D.Baker$^2$, M.Ballintijn$^4$, D.S.Barton$^2$, R.R.Betts$^6$,
R.Bindel$^7$, W.Busza$^4$, Z.Chai$^2$, V.Chetluru$^6$,
E.Garc\'{\i}a$^6$, T.Gburek$^3$, K.Gulbrandsen$^4$,
J.Hamblen$^8$,I.Harnarine$^6$, C.Henderson$^4$,
D.J.Hofman$^6$,~R.S.Hollis$^6$, R.Ho\l
y\'{n}ski$^3$,~B.Holzman$^2$,~A.Iordanova$^6$,~J.L.Kane$^4$,
P.Kulinich$^4$,vC.M.Kuo$^5$,~W.Li$^4$,~W.T.Lin$^5$,~C.Loizides$^4$,
S.Manly$^8$,~A.C.Mignerey$^7$,~R.Nouicer$^2$, A.Olszewski$^3$,
R.Pak$^2$, C.Reed$^4$, E.Richardson$^7$, C.Roland$^4$, G.Roland$^4$,
J.Sagerer$^6$, I.Sedykh$^2$, C.E.Smith$^6$,
M.A.Stankiewicz$^2$,P.Steinberg$^2$, G.S.F.Stephans$^4$,
A.Sukhanov$^2$, A.Szostak$^2$, M.B.Tonjes$^7$, A.Trzupek$^3$,
G.J.van~Nieuwenhuizen$^4$, S.S.Vaurynovich$^4$, R.Verdier$^4$,
G.I.Veres$^4$, P.Walters$^8$,~E.Wenger$^4$, D.Willhelm$^7$,
F.L.H.Wolfs$^8$, B.Wosiek$^3$, K.Wo\'{z}niak$^3$, S.Wyngaardt$^2$,
B.Wys\l ouch$^4$. $^1$~Argonne National Laboratory, Argonne, IL
60439-4843, USA. $^2$~Brookhaven National Laboratory, Upton, NY
11973-5000, USA.  $^3$~Institute of Nuclear Physics PAN, Krak\'{o}w,
Poland. $^4$~Massachusetts Institute of Technology, Cambridge, MA
02139-4307, USA. $^5$~National Central University, Chung-Li, Taiwan.
$^6$~University of Illinois at Chicago, IL 60607-7059, USA.
$^7$~University of Maryland, MD 20742,~USA.  $^8$~University of
Rochester, Rochester, NY 14627,~USA.}
\section{INTRODUCTION}
The azimuthal correlations of produced particles have proven to be a
sensitive measure of the initial conditions and subsequent dynamics in
relativistic heavy ion collisions.  The elliptic flow
$\mathrm{v_{2}}$, as inferred from the angular distribution of
particles with respect to the reaction plane, provides important
constraints on hydrodynamical descriptions about the evolution of the
collision~\cite{Hirano}.  The large pseudorapidity coverage (${\rm
|\eta|\le 5.4}$) and the near symmetric azimuthal acceptance for
charged hadrons of the PHOBOS detector at RHIC make it excellent for
the investigation of the systematics of the flow measurements, as a
function of energy, system size, centrality and pseudorapidity.  This
contribution will concentrate on $\mathrm{v_{2}}$ measurements for the
Cu+Cu collisions at \mbox{$\mathrm{\sqrt{s_{NN}}}$} = 62.4 and 200 GeV
and a comparison to previously reported Au+Au results.  Details of the
PHOBOS detector and the experimental technique used to extract the
flow can be found in references {\cite{1Nim, 2JoshsFlow,3CarlasFlow}}.
\section{RESULTS AND CONCLUSIONS}
The pseudorapidity distributions of the elliptic flow,
$\mathrm{v_{2}(\eta)}$, for the Au+Au and Cu+Cu collisions over broad
range of center-of-mass energies and for a centrality of 0-40\% are
shown in Fig. 1. The triangular shape first observed for Au+Au
collisions~\cite{2JoshsFlow} is also apparent in the Cu+Cu
data~\cite{4SteveQM}.  The measurements of $\mathrm{v_{2}}$ for Cu+Cu collisions is only about
20\% lower than that for Au+Au results, even though the Cu+Cu system size is
about a third of that of Au+Au.
\begin{figure}[]
  \centerline{
      \includegraphics[width=1.0\textwidth]{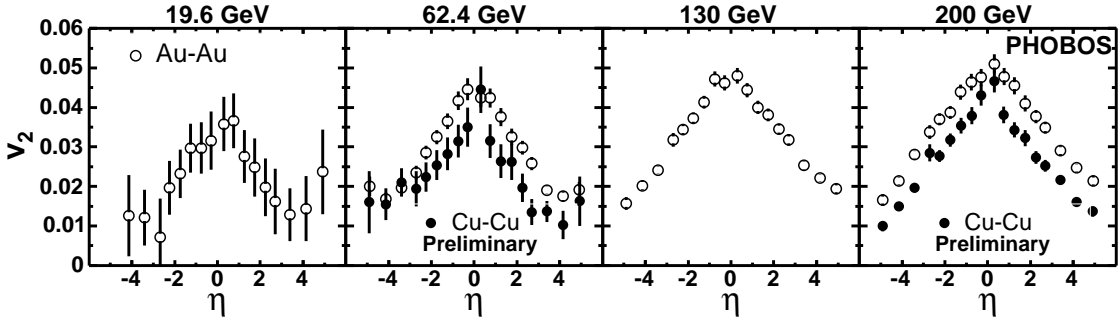}
  }
  \caption{ Measured $\mathrm{v_{2}(\eta)}$ from Au+Au and Cu+Cu
  collisions at RHIC energies and for the centrality range of 0-40\%.
  Only the 1$\mathrm{\sigma}$ statistical error bars are shown for
  clarity. The full systematic errors can be found in Ref.~
  {\cite{2JoshsFlow}} for the Au+Au data and Ref.~{\cite{4SteveQM}} for the
  Cu+Cu data.}
  \label{fig_v2_eta}
\end{figure}
The measured pseudorapidity density of charged hadrons,
\mbox{$\mathrm{\frac{dN_{ch}}{d\eta}}$}, for Cu+Cu and Au+Au
collisions at \mbox{$\mathrm{\sqrt{s_{NN}}}$} = 200 GeV, is
essentially the same at a given $\mathrm{N_{part}}$
{\cite{5GuntherQM}}, but the elliptic flow is different due to the
difference in system size. While $\mathrm{\langle N_{part} \rangle}$
of 100 corresponds to a 3-6\% centrality selection for the Cu+Cu
system, $\mathrm{\langle N_{part} \rangle}$ of 99 corresponds to only
35-40\% central for Au+Au.  This implies different initial
geometrical overlaps for the two systems with the same energy density
(as reflected in the near identical
\mbox{$\mathrm{\frac{dN_{ch}}{d\eta}}$}). The measured
$\mathrm{v_{2}}$ for Au+Au and Cu+Cu collisions at the same energy,
200 GeV, expressed as a function of $\mathrm{N_{part}}$ is shown in
Fig. 2.
\begin{center}

\begin{figure}[]
\centering      \includegraphics[height=.23\textheight]
		      {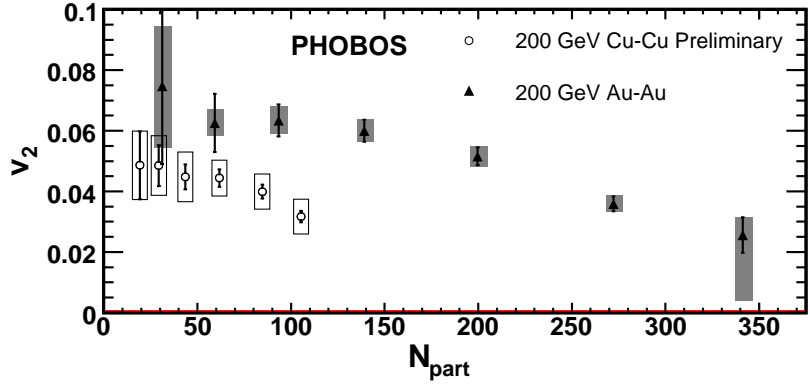} 
      \caption
      {The elliptic flow $\mathrm{v_{2}}$ as a function of
      $\mathrm{N_{part}}$ measured at midrapidity
      (\mbox{$\mathrm{|\eta|<1}$}), for Au+Au and Cu+Cu at
      \mbox{$\mathrm{\sqrt{s_{NN}}}$} = 200 GeV. The Au+Au and Cu+Cu data
      are from Ref.~{\cite{3CarlasFlow}} and {\cite{4SteveQM}}, respectively.
      The error bars represent the 1$\mathrm{\sigma}$ statistical errors
      and the boxes are 90\% C.L. systematic errors. }
\end{figure}
\end{center}

\begin{figure}[ht]
  \centerline{
      \includegraphics[width=1.0\textwidth]{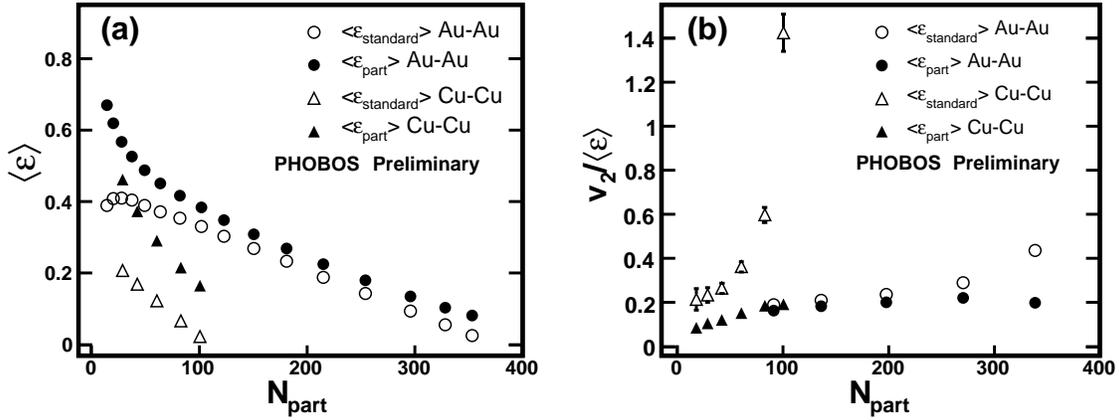}
  }
  \caption{Panel a): Mean standard and participant eccentricities, calculated using a Glauber model and panel b): elliptic flow $\mathrm{v_{2}}$ normalized by the two eccentricities, for Au+Au and Cu+Cu collisions at  \mbox{$\mathrm{\sqrt{s_{NN}}}$} = 200 GeV, as a function of the number of participants $\mathrm{N_{part}}$.  Only the 1$\mathrm{\sigma}$ statistical errors of $\mathrm{v_{2}}$ are reflected in the error bars in panel b). }
  \label{fig_v2_ecc}
\end{figure}

  This discrepancy in elliptic flow for the two systems can be
  accounted for if $\mathrm{v_{2}}$ is normalized by the eccentricity
  of the system.  Two definitions of eccentricity have been used,
  based on a simple Glauber model {\cite{4SteveQM, 5GuntherQM}}.  The
  standard eccentricity, $\mathrm{\epsilon_{standard}}$, is defined in
  the frame of the original impact parameter; a participant
  eccentricity, $\mathrm{\epsilon_{part}}$, is obtained from the
  geometry of the participants that define $\mathrm{N_{part}}$, and is
  influenced by fluctuations in the participant positions. This is
  more important for the much smaller Cu+Cu system than for Au+Au, as
  seen in Fig. 3(a), where the mean eccentricities derived from the
  two approaches are compared.  The $\mathrm{v_{2}}$ normalized by the
  two eccentricities is shown in Fig. 3(b).  While
  $\mathrm{v_{2}}$/$\mathrm{\langle\epsilon_{part}\rangle}$ is nearly identical for
  the Cu+Cu and Au+Au systems,
  $\mathrm{v_{2}}$/$\mathrm{\langle\epsilon_{standard}\rangle}$ gives
  unrealistically large values for the most central events.  It is
  clear that a better understanding of the eccentricity relevant to
  the reaction dynamics is needed in order to meaningfully compare
  systems of such different sizes as Cu+Cu and Au+Au.

\begin{theacknowledgments}
This work was partially supported by U.S. DOE grants
DE-AC02-98CH10886, DE-FG02-93ER40802, DE-FC02-94ER40818,
DE-FG02-94ER40865, DE-FG02-99ER41099, and W-31-109-ENG-38, by U.S. NSF
grants 9603486, 0072204, 0245011, by Polish KBN grant
1-P03B-062-27(2004-2007), by NSC of Taiwan Contract NSC
89-2112-M-008-024, and by Hungarian OTKA grant (F 049823).
\end{theacknowledgments}



 \bibliographystyle{aipprocl} 

\begin{thebibliography}{5}

{\bibitem{Hirano}      T. Hirano, \emph{Proceedings of PANIC} (2005), arXiv:nucl-th/0601006.	}
{\bibitem{1Nim}        B.~B.~Back {\it et al.}, Nucl. Instrum. Methods A {\bf 499}, 603 (2003).	}
{\bibitem{2JoshsFlow}  B.~B.~Back {\it et al.}, Phys. Rev. Lett. {\bf 94}, 122303 (2005).		}
{\bibitem{3CarlasFlow} B.~B.~Back {\it et al.}, Phys. Rev. C {\bf 72} 051901R (2005).			}
{\bibitem{4SteveQM}    S.~Manly (for the PHOBOS Collaboration), \emph{Proceedings of QM} (2005), arXiv:nucl-ex/0510031.		}
{\bibitem{5GuntherQM}  G.~Roland (for the PHOBOS Collaboration), \emph{Proceedings of QM} (2005), arXiv:nucl-ex/0510042.	}

\end{thebibliography}

\end{document}